\documentclass[aps,prl,twocolumn,showpacs]{revtex4}
\usepackage{amsmath}
\usepackage{epsfig}
\usepackage{graphicx}

\def\dfrac#1#2{{\displaystyle{#1\over#2}}}
\def\vec#1{{\bf#1}}
\def\eq#1{Eq.~(\ref{#1})}
\def\mb#1{\mbox{\boldmath$#1$}}
\def\fig#1{Fig.~\ref{#1}}

\def\be{\begin{equation}}
\def\ee{\end{equation}}
\def\bea{\begin{eqnarray}}
\def\eea{\end{eqnarray}}

\begin{document}

\title{DC Spin Current Generation in a Rashba-type Quantum Channel}
\author{L. Y. Wang,$^{1}$ C. S. Tang,$^{2}$ and C. S. Chu$^{1}$}
\affiliation{$^{1}$Department of Electrophysics, National Chiao
Tung University, Hsinchu 30010, Taiwan\\
$^{2}$Physics Division, National Center for Theoretical Sciences,
P.O. Box 2-131, Hsinchu 30013, Taiwan}
\date{\today }

\begin{abstract}
\ We propose and demonstrate theoretically that resonant inelastic
scattering (RIS) can play an important role in dc spin current
generation. The RIS makes it possible to generate dc spin current
via a simple gate configuration: a single finger-gate that locates
atop and orients transversely to a quantum channel in the presence
of Rashba spin-orbit interaction. The ac biased finger-gate gives
rise to a time-variation in the Rashba coupling parameter, which
causes spin-resolved RIS, and subsequently contributes to the dc
spin current. The spin current depends on both the static and the
dynamic parts in the Rashba coupling parameter, $\alpha_0$ and
$\alpha_1$, respectively, and is proportional to $\alpha_0
\alpha_1^2$. The proposed gate configuration has the added
advantage that no dc charge current is generated. Our study also
shows that the spin current generation can be enhanced
significantly in a double finger-gate configuration.
\end{abstract}





\maketitle

Spintronics is important in both application and fundamental
arenas~\cite{Loss02,Wolf01}. A recent key issue of great interest
is the generation of dc spin current (SC) without charge current.
Various dc SC generation schemes have been proposed, involving
static magnetic field~\cite{Mucc02, Qing03}, ferromagnetic
material~\cite{Brat02}, or ac magnetic field~\cite{Ping03}. More
recently, Rashba-type spin-orbit interaction in 2DEG~\cite{Rash84,
Nitta97} has inspired attractive proposals for nonmagnetic dc SC
generation~\cite{Pras03, Gove03, Malsh03}. Of these recent
proposals, including a time-modulated quantum dot with a static
spin-orbit coupling~\cite{Pras03}, and time-modulations of a
barrier and the spin-orbit coupling parameter in two spatially
separated regions~\cite{Gove03}, the working principle is
basically adiabatic quantum pumping. Hence simultaneous generation
of both dc spin and charge current is the norm. The condition of
zero dc charge current, however, is met only for some judicious
choices for the values of the system parameters.

It is known, on the other hand, that quantum transport in narrow
channel exhibits RIS features when it is acted upon by a spatially
localized time-modulated potential ~\cite{Bagw92, Tang96}. This
RIS is coherent inelastic scattering, but with resonance at work,
when the traversing electrons can make transitions to their
subband threshold by emitting $m\hbar\Omega$~\cite{Bagw92,
Tang96}. Should this RIS become spin-resolved in a Rashba-type
quantum channel (RQC), of which its Rashba coupling parameter is
time-modulated locally, we will have a simpler route to the
nonmagnetic generation of dc SC. Thus we opt to study, in this
Letter, the RIS features in a RQC. As is required by a study on
the RIS features, our study goes beyond the adiabatic regime.

\begin{figure}[b]
\includegraphics[width=.4 \textwidth]{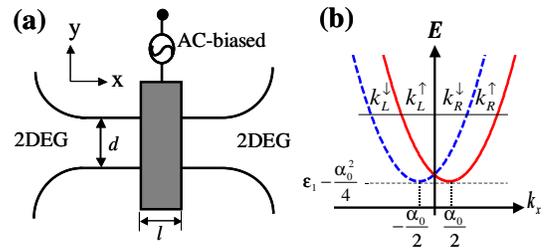}
\caption{(a) Top-view schematic illustration of the RQC. The
ac-biased FG, of width $l$, is indicated by the grey area; (b) the
electron dispersion relation of an unperturbed RQC.}\label{sys}
\end{figure}

The system configuration considered is based on a RQC that forms
out of a 2DEG in an asymmetric quantum well by the split-gate
technique. As is depicted in \fig{sys}(a), a finger gate (FG) is
positioned above while it is separated from the RQC by an
insulating layer. A local time-variation in the Rashba coupling
parameter $\alpha(\mathbf{r},t)$ can be induced by ac biasing the
FG~\cite{Gove03, Malsh03}. The Hamiltonian is given by
$\mathcal{H} = p^2/2m + \mathcal{H}_{\mathrm{so}}(\mathbf{r},t) +
V_{\rm c}(y)$ where the Rashba term
\begin{equation}
\mathcal{H}_{\mathrm{so}}(\mathbf{r},t) =
\mathbf{M}\cdot\frac{1}{2}\left[ {\alpha
\left(\mathbf{r},t\right)\mbox{\boldmath$p$} +
\mbox{\boldmath$p$}\,\alpha\left(\mathbf{r},t\right)} \right]\, .
\end{equation}
Here $\mathbf{M}=\hat{\mb {z}}\times\mbox{\boldmath$\sigma$}$,
$\hat{\mb{z}}$ is normal to the 2DEG, $\mbox{\boldmath$\sigma$}$
is the vector of Pauli spin matrices, and $V_{\rm c}(y)$ is the
confinement potential. The unperturbed Rashba coupling parameter
$\alpha(\mathbf{r},t)$ is $\alpha_0$ throughout the RQC, but when
perturbed by the ac biased FG, it becomes $\alpha_0 +
\alpha_1\cos\Omega t$ in the region underneath the FG. The
Dresselhaus term is neglected for the case of a narrow gap
semiconductor system~\cite{Lomm88}.

For a clear demonstration of the pumping mechanism, the
unperturbed RQC we considered is narrow so that its subband energy
spacing is much greater than the subband mixing due to the Rashba
interaction. As such, the unperturbed Hamiltonian, in its
dimensionless form, is $\mathcal{H}_{0} = -\mathbf{\nabla}^2 +
\alpha_0 \sigma_y
 (i\partial / \partial x) + V_\mathrm{c}(y)$. Appropriate
units have been used such that all physical quantities presented
here, and henceforth, are dimensionless~\cite{Tang96}. In
particular, $\alpha$ is in unit of $v_{{\rm F}}/2$, $v_{{\rm F}}$
denoting the Fermi velocity, and spin in unit of $\hbar/2$. The
right-going (R) eigenstate of $\mathcal{H}_{0}$, in the $n$-th
subband, is $\phi_n (y)\,\psi_{n}^\sigma(x)$, where
$\psi_{n}^\sigma(x) = \exp \left[ik_{n,R}^\sigma x\right]
\chi_\sigma$. The wavevector $k_{n,R}^\sigma = \sqrt{\mu_n} +
\eta_\sigma \alpha_0 /2$ while $\eta_\sigma=\pm 1$ denotes the
eigenvalue of $\chi_\sigma$ to the operator $\sigma_y$. $\mu_n$ is
the energy measured from the $n$-th subband threshold such that
the energy of the eigentstate is $E = \mu_n + \varepsilon_n -
\alpha_0^2/4$, for $\varepsilon_n = (n \pi/d)^2$. This dispersion
relation is shown in \fig{sys}(b). It is of import to note, for
later reference, that right-going electrons have $|
k_{\uparrow}^{R}| \,> \,| k_{\downarrow}^{R}|$, and that at the
subband threshold $k_{\uparrow(\downarrow)}^{R} \,=\,
k_{\uparrow(\downarrow)}^{L}$.

In the ac-biased region,
$\mathcal{H}=\mathcal{H}_{x}+\mathcal{H}_{y}$, the transverse part
$\mathcal{H}_y = -{\partial}^2/\partial{y}^2 + V_\mathrm{c}(y)$,
and the longitudinal part
\begin{equation}
\label{Hx}\mathcal{H}_{x} \left(t\right) =
\left(-i\frac{\partial}{\partial x} +
\frac{\alpha(x,t)}{2}\mathbf{M}\cdot \hat{\mb{x}} \right)^2 -
\frac{1}{4}\alpha\left(x,t\right)^2.
\end{equation}
The form of \eq{Hx} suggests an effective vector potential,
$\mathbf{A}(t) = \frac{1}{2}\alpha(x,t)\mathbf{M}\cdot
\hat{\mb{x}}$, which depends on the spin and gives rise to a
spin-resolved driving electric field $\mathbf{E} = -
\partial\mathbf{A} / \partial t$. However, in
$\mathcal{H}_{x}$, the $A^{2}$ term does not depend on
$\mbox{\boldmath$\sigma$}$, while for the term linear in $\vec{A}$,
$\vec{A}\,\chi_\sigma=-\frac{1}{2}\eta_{\sigma}\,\alpha(x,t)\,\chi_\sigma$
gives rise only to a trivial spin dependence, which can be easily
removed by a shift in the origin of time for the case of an
oscillatory $\alpha(x,t)$. Yet it turns out that the full term
linear in $\vec{A}$, given by $-i\frac{\partial}{\partial
x}\,\hat{x}\cdot \vec{A}$, manages to give rise to nontrivial
spin-resolved transmissions. In a perturbative sense, this term
becomes $k_{\uparrow(\downarrow)}^{R}\,A_{x}$, for the case of a
right-going electron incident upon a spatially uniform $\alpha(t)$.
This renders the effective longitudinal driving field to become spin
dependent. As a consequence, the difference in the current
transmissions, for spin up and spin down cases, is proportional to
$\alpha_{0}$, and the difference is found to be amplified by RIS. In
a RQC that has zero source-drain bias, the spin-resolved current
transmission leads readily to dc spin current, but it cannot lead to
dc charge current if the RQC is symmetric with respect to its source
and drain.

An alternate way to understand the origin of the spin-resolved
current transmission is presented in the following.
Performing a unitary transformation $\Psi _\sigma (x,t) = \exp
\left[ {(i\eta_\sigma /2)\int^{x}_{-l/2}{\alpha \left( {x',t}
\right)dx'} } \right]\psi _\sigma (x,t)$, the Schr\"odinger
equation, \eq{Hx}, becomes
\begin{equation} \label{SE}
\left[ { - \frac{\partial^2}{\partial x^2} + U_1 \left( t \right) +
U_{2}^{\sigma}\left(t\right)} \right]\psi _\sigma \left( {x,t}
\right) = i\frac{\partial } {{\partial t}}\psi _\sigma \left( {x,t}
\right),
\end{equation}
of which the two time-dependent potentials are
$U_1(t)=-\alpha(x,t)^2/4$, and $U_{2}^{\sigma}(t)= \left(\Omega\,
\alpha _1/2\right)\left(x + l/2\right)$ $\cos \left(\Omega
t+\,\eta_\sigma\pi/2\right)$. Even though only $U_{2}^{\sigma}$
depends on spin, both the term in $U_1(t)$ that oscillates with
frequency $\Omega$ and $U_{2}^{\sigma}$ together constitute a pair
of quantum pumping potential that pump SC. This is our major
finding in this work: that spin pumping nature is build-in even in
a single FG configuration. Its origin is rooted in the intricate
Rashab spin dynamics.

The expression for the pumped SC, in the absence of the source-drain
bias, is obtained from the SC density operator
\begin{equation}\label{Jxy}
\hat{J}_x^y  = i\left[ \frac{\partial \Psi^\dag_\sigma}{\partial x}
\sigma _y \Psi_\sigma - \text{H. c.}\right] + {\frac{\alpha}{2}}
\Psi^\dag_\sigma  \left\{ {\sigma _y ,\mathbf{M}}
\right\}_x\Psi_\sigma.
\end{equation}
The SC conservation is maintained due to the supression of subband
mixing in a RQC, and the associated spin-flip mechanism. By taking
the time-average of the transmitted and the incident SC in
\eq{Jxy}, and their ratio, we obtain the spin-resolved current
transmission $T_{\beta\alpha}^{\sigma}$, where $\alpha$, $\beta$,
are, respectively, the incident and the transmitting lead. Summing
over all possible incident states from both reservoirs $R$ and
$L$, the net SC is given by
\begin{eqnarray}
 I^s &=& I^{\uparrow} + I^{\downarrow} \nonumber \\
&=&  \int dE\,f(E)\,\left[ \left(T_{RL}^{\uparrow}-
T_{LR}^{\uparrow} \right) + \left( T_{LR}^{\downarrow}-
T_{RL}^{\downarrow} \right)\right], \label{I}
\end{eqnarray}
where $f(E)$ is the Fermi-Dirac distribution, and $T^\sigma_{RL} =
\sum_n \sum_{m (\mu_n^m>0)} T^{m,\sigma}_{n,RL}$. The transmission
coefficient $T^{m,\sigma}_{n,RL} = \left| t_{n,RL}^{m,\sigma}
\right|^2 \sqrt{\mu_n^m/\mu_n}$ denotes the current transmission
that an electron incident from terminal $L$ in the spin channel
$\sigma$, subband $n$, energy $E$, is scattered into terminal $R$,
sideband $m$, with kinetic energy $\mu^{m}_{n}=\mu_{n}+m\Omega$.
The net charge current is simply given by $I^q = I^{\uparrow} -
I^{\downarrow}$. In a symmetric time-dependent FG configuration,
we have $T_{LR}^{\sigma} = T_{RL}^{-\sigma}$, so that the net spin
current is $I^s = 2 \int dE \,f(E)\, \left( T_{RL}^{\uparrow}-
T_{RL}^{\downarrow} \right)$ and the net charge current is
identically zero.

Before we demonstrate numerically the robustness of the SC pumping,
it is worthwhile to first look at the weak-pumping regime, which
affords us analytical results. To this end, the scattering
amplitudes are restricted to include only up to the sidebands $m=\pm
1$, that is, up to first power in $\alpha_1$. For an electron
incident with wave vector $k_{n,R}^\sigma$ from terminal $L$, the
spin-resolved reflection amplitude is obtained to be
\begin{widetext}
\begin{equation}
r_{n,LL}^{m,\sigma} = \eta _\sigma \left(\dfrac{ \alpha
_1}{2}\right) \frac{{\left[  e^{i\left( {k_{n,R}^{\sigma} -
k_{n,L}^{m,\sigma}} \right)l}-1 \right]\left[
\dfrac{1}{\Omega}{k_{n,R}^{\sigma} \left( {k_{n,R}^{\sigma}  -
k_{n,R}^{m,\sigma} } \right) + \dfrac{m}{2}}
\right]}}{k_{n,R}^{m,\sigma}  - k_{n,L}^{m,\sigma}}
\end{equation}
\end{widetext}
for $m=\pm 1$. The reflection amplitude $r_{n,LL}^{0,\sigma}$ is
of order $\alpha_1^2$, and is negligible in our weak-pumping
approximation. Here the wave vector $k_{n,R(L)}^{m,\sigma} = \pm
(\mu_n^m)^{1/2} + \eta_\sigma \alpha_0/2$, with upper (lower) sign
corresponds to the right- (left-) moving electron in the $n$th
subband, $m$th sideband, and with kinetic energy $\mu_n^m$. It is
obvious then that differences between wave vectors of different
sideband indices, but of the same spin state, are spin
independent. Hence the spin dependence of $r_{n,LL}^{\pm
1,\sigma}$ arises solely from the $k^\sigma_{n,R}$ in the
numerator that does not involve in a wave vector difference.
Furthermore, this spin dependence is associated with $\alpha_{0}$.
It is not unexpected then that the SC is proportional to
$\alpha_{0}$. The SC is related to the current transmission which,
within the aforementioned approximation, is given by $
T^\sigma_{RL} \approx 1 - \sum_n \left[ R^{1,\sigma}_{n,LL} +
R^{-1,\sigma}_{n,LL} \right]$, where $R^{m,\sigma}_{n,LL} =
\left|r_{n}^{m,\sigma}\right|^2 \sqrt{\mu_n^m}/\sqrt{\mu_n}$. From
\eq{I}, the energy derivative of the zero temperature SC is given
by $\partial I^{s}/\partial E = 2\Delta T_{RL} = 2(T_{RL}^\uparrow
- T_{RL}^\downarrow)$ from which its explicit expression is given
by
\begin{widetext}
\begin{equation}
\dfrac{\partial I^{s}}{\partial E} =
\dfrac{1}{2}\alpha_0\alpha_1^2 \sum\limits_{n} \sum\limits_{
\begin{subarray}{l}
  m =  \pm 1 \\
             \\
  (\mu _n^m  > 0)
\end{subarray}}
\frac{\left[ 1 - \cos \left( \left( \sqrt{\mu_n} + \sqrt
{\mu_n^m} \right)l \right) \right] \left[\left( \dfrac{1}{4}
\right)^2 - \left( \dfrac{1}{\Omega^2}\left( \mu_n + \sqrt
{\mu_n\mu_n^m}\right) +\dfrac{m}{4} \right)^{2}\right]} {\mu_n
\sqrt{\mu_n^m} }. \label{dIdE}
\end{equation}
\end{widetext}
That this expression diverges when $\mu_{n}^{m}=0$, for $m<0$,
exhibits the RIS feature unambiguously.

\begin{figure}[th]
\includegraphics[width=.42 \textwidth]{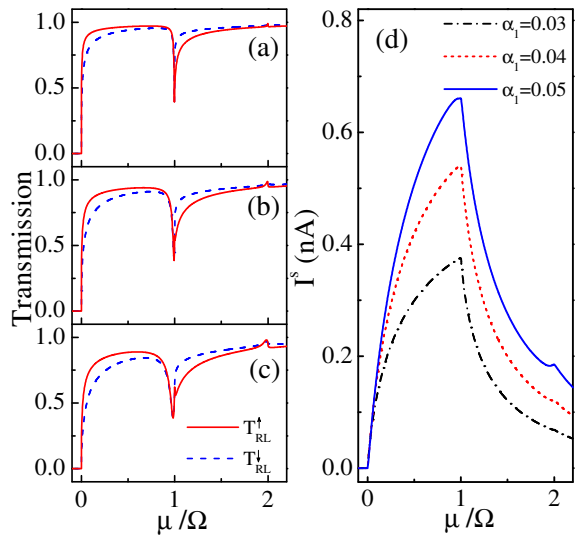}
\caption{Spin-resolved current transmissions $T_{RL}^{\uparrow}$
(solid) and $T_{RL}^{\downarrow}$ (dashed) versus the incident
energy $\mu / \Omega$.  Parameters $N=1$, $\alpha_0=0.13$,
$\Omega=0.002$, $l=20$, and with $\alpha_1$ = (a) $0.03$, (b)
$0.04$, and (c) $0.05$. The corresponding dc SC is plotted in (d).
\label{N1} }
\end{figure}

In the following, we present results obtained from solving the
time-dependent spin-orbit scattering exactly, in the numerical
sense~\cite{Wang04}. Physical parameters are chosen to be
consistent with the InGaAs-InAlAs based narrow gap
heterostructures such that the electron density $n_e=1 \times
10^{12}$ cm$^{-2}$, effective mass $m^* = 0.04 m_0$, and $\alpha_0
= 0.13$ ($\hbar\alpha_0 = 3\times10^{ - 11}$ eV m)~\cite{Nitta97}.
Accordingly, the length unit $l^* = 4.0$ nm, and the energy unit
$E^* = 59$ meV.

For the case of one FG ($N=1$), the energy dependence of the
spin-resolved transmission $T_{RL}^{\sigma}$ is plotted in
Figs.\,\ref{N1} (a)-(c), and that of the corresponding dc SC is
plotted in Fig.\,\ref{N1} (d). The FG width $l$ = 20 (80 nm),
driving frequency $\Omega = 0.002$ ($\nu = \Omega /2\pi \approx 28$
GHz), and energy $\mu = E - \varepsilon_1$. Dip features in
$T_{RL}^{\sigma}$ at $\mu /\Omega= 1$ are the QBS features, where
electrons undergo coherent inelastic scattering to a QBS just
beneath its subband bottom~\cite{Bagw92}. Higer order QBS features
at $\mu/\Omega = 2$ are barely shown by the small peaks. Of
particular interest is the change in sign in the transmission
difference $\Delta T_{RL} = T_{RL}^{\uparrow} - T_{RL}^{\downarrow}$
across the dip structures, namely, $\Delta T_{RL}(\mu = \Omega^{-})
> 0$ while $\Delta T_{RL}(\mu = \Omega^{+}) < 0$. This leads to a
nonzero dc SC, and that it peaks at $\mu/\Omega = 1$, as is
exhibited in \fig{N1}(d). It is also shown that the dc SC increases
with the oscillating amplitude $\alpha_1$ of the ac-biased gate
voltage.

\begin{figure}[th]
\includegraphics[width=.42 \textwidth,angle=0]{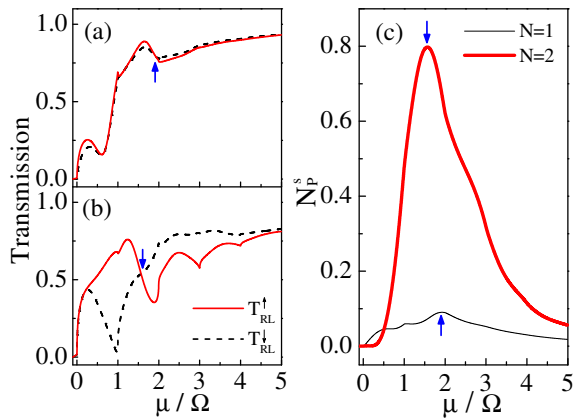}
\caption{Current transmission versus $\mu/\Omega$ for $N=$ (a) 1,
and (b) 2. Pumped spins per cycle are plotted in (c) for $N=1$
(thick curve) and $N=2$ (thin curve) with driving frequency
$\Omega=0.001$. Other parameters are the same as in \fig{N1}.}
\label{N12}
\end{figure}

The possibility of nonlinear enhancement in the dc SC by two FGs
($N=2$) is presented in Figs.\,\ref{N12} (a)-(c). The driving
frequency is chosen to be $\Omega=0.001$ ($\nu \approx {14}$ GHz),
and the FG width $l=22$ ($\simeq$ 88 nm). For comparison, the
$N=1$ FG transmissions are plotted along side with that of the
$N=2$ FG case, in Figs.\,\ref{N12} (a), and (b), respectively. The
corresponding dc SC, expressed in terms of pumped spin per cycle
$N_P^s=(2\pi/\Omega)|I_R^s|$, is shown in \fig{N12}(c). The
pumping is optimized by a choice of the FG separation, with the
edge to edge separation $\Delta l = 22$. That nonlinear effects
are significant is supported by the appearance of up to the
fourth-sideband QBS dip structures in Fig.\,~\ref{N12}(b). As
indicated by arrows, the pumped spin per cycle peaks at
$\mu/\Omega \simeq 1.57$ ($1.92$), and with peak value 0.8 (0.1)
for the case of $N=2$ ($N=1$) FG. The enhancement in $N_P^s$ is
nonlinear, far greater than doubling the $N_P^s$ of $N=1$ FG.

\begin{figure}[th]
\includegraphics[width=.44 \textwidth,angle=0]{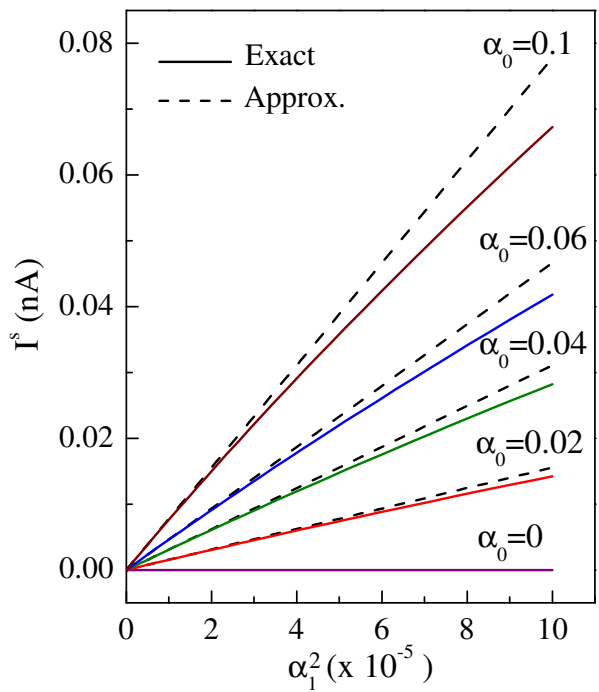}
\caption{Pumped dc SC versus $\alpha_1^2$ for various $\alpha_0$
values. Other parameters are $\Omega=0.01$ and $l=20$.}
\label{alpha}
\end{figure}

In \fig{alpha}, we present the dependence of the dc SC peak values
on the pumping parameters $\alpha_{0}$ and $\alpha_{1}$, and for
the case of one FG. The numerical results, depicted by solid
curves, coincide nicely, in the small $\alpha_{1}$ regime, with
the one sideband approximation results, depicted by broken curves,
and calculated according to Eq. (7). This confirms that the dc SC
is proportional to $\alpha_0\alpha_1^2$ in the weak pumping
regime. Moreover, deviation of the numerical results from the weak
pumping behavior sets in at smaller $\alpha_{1}$ values when
$\alpha_{0}$ increases. Typical degree of deviation can be
inferred from the case of $\alpha_0 = 0.1$ and $\alpha_1 = 0.01$,
where the deviation of dc SC $\Delta I^s =
|I_{\mathrm{numerical}}^s - I_{\mathrm{app}}^s|/
I_{\mathrm{numerical}}^s \simeq 0.16$. We also find that, as
$\Omega$ decreases, the degree of deviation increases, indicating
the need to include more sidebands for the description of the
time-dependent quantum scattering.

In conclusion, a nonmagnetic way of generating dc SC has been
established. The proposed configuration, a Rashba-type quantum
channel driven by an ac biasing finger gate, is relatively simple
and is within reach of recent fabrication capability. The nature of
the spin pumping is studied in detail and its pumping mechanism
understood. Resonant inelastic process is found to be a major factor
that contributes to the robustness of the spin pumping. The coherent
nature of the pumping supports further enhancement of the spin
pumping by invoking configuration consisting of more than one finger
gates.

\begin{acknowledgements}
The authors acknowledge valuable discussions with A. G.
Mal'shukov. This work was funded by the National Science Council
of ROC under Grant Nos. NSC92-2112-M-009-035,
NSC92-2120-M-009-010, NSC93-2112-M-009-036 and
NSC93-2119-M-007-002 (NCTS).
\end{acknowledgements}


\begin{thebibliography}{99}

\bibitem{Loss02} \textit{Semiconductor Spintronics and Quantum
Computation}, edited by D.D. Awschalom, N. Samarth, and D. Loss
(Springer-Verlag, Berlin, 2002).

\bibitem{Wolf01} S.A. Wolf \textit{et al.},
Science \textbf{294}, 1488 (2001); Y. Kato \textit{et al.},
\textit{ibid.} \textbf{299}, 1201 (2003); S. Murakami \textit{et
al.}, \textit{ibid.} \textbf{301}, 1348 (2003).

\bibitem{Mucc02} E. R. Mucciolo, C. Chamon, and C. M. Marcus,
Phys. Rev. Lett. \textbf{89}, 146802 (2002). Experimental
realization was reported by S. K. Watson, R. M. Potok, C. M.
Marcus, and V. Umansky in Phys. Rev. Lett. \textbf{91}, 258301
(2003).

\bibitem{Qing03} Q. F. Sun, H. Guo, and J. Wang, Phys. Rev. Lett.
\textbf{90}, 258301 (2003).

\bibitem{Brat02} A. Brataas, Y. Tserkovnyak, G. E. W. Bauer, and
B. I. Halperin, Phys. Rev. B \textbf{66} 60404 (2002).

\bibitem{Ping03} P. Zhang, Q. K. Xue, and X. C. Xie, Phys. Rev.
Lett. \textbf{91}, 196602 (2003).

\bibitem{Rash84} Y. A. Bychkov and E. I. Rashba, J. Phys. C
\textbf{17}, 6039 (1984).

\bibitem{Nitta97} J. Nitta \textit{et al.}, Phys. Rev. Lett.
\textbf{78}, 1335 (1997); D. Grundler, \textit{ibid.} \textbf{84},
6074 (2000).

\bibitem{Pras03} P. Sharma and P. W. Brouwer, Phys. Rev. Lett.
\textbf{91}, 166801 (2003).

\bibitem{Gove03} M. Governale, F. Taddei, and R. Fazio, Phys. Rev.
B \textbf{68}, 155324 (2003).

\bibitem{Malsh03} A. G. Mal'shukov, C. S. Tang, C. S. Chu, and K. A. Chao,
Phys. Rev. B \textbf{68}, 23 3307 (2003).

\bibitem{Bagw92} P. F. Bagwell and R. K. Lake, Phys. Rev. B \textbf{46},
15329 (1992).

\bibitem{Tang96} C. S. Tang and C. S. Chu, Phys. Rev. B
\textbf{53}, 4838 (1996).

\bibitem{Lomm88} G. Lommer, F. Malcher, and U. R$\ddot{o}$ssler,
Phys. Rev. Lett. \textbf{60}, 728 (1988).

\bibitem{Wang04} L. Y. Wang, C. S. Tang, and C. S. Chu
(unpublished).





\end{thebibliography}
\end{document}